\documentclass[12pt]{article}
\usepackage{amsmath,amssymb,amsthm}
\usepackage{mathtools}
\usepackage{graphics,epsfig}
\usepackage{hyperref}
\usepackage{natbib}
\usepackage{color}
\usepackage{graphicx}
\usepackage{caption}
\usepackage{subcaption}
\usepackage{float}
\usepackage{mathrsfs}
\usepackage{multirow}
\usepackage{comment}
\usepackage{setspace}
\usepackage{bbm}
\usepackage{bm}
\usepackage{amsfonts}
\usepackage{xcolor}
\usepackage{pslatex}
\usepackage{setspace}
\usepackage{bbm}
 \usepackage{booktabs}


\begin{document}

  \title{\bfseries Modeling Age-Adjusted Mortality in the United States }

  \author{
    Brandon Dunbar$^{1}$ \and Paramahansa Pramanik$^{2,3}$\and Haley Kate Robinson $^{4}$
}

\date{
    \small
    $^{1}$Department of Mathematics and Statistics, University of South Alabama, Mobile, AL 36688, United States.\\
    \texttt{bjd2225@jagmail.southalabama.edu} \\
    \vspace{0.5em}
    $^{2}$Department of Mathematics and Statistics, University of South Alabama, Mobile, AL 36688, United States.\\
    $^{3}$Corresponding author, \texttt{ppramanik@southalabama.edu}\\
    $^{4}$ Department of Biomedical Sciences, University of South Alabama,
    Mobile, AL 36688,  United States. \texttt{hkr2322@jagmail.southalabama.edu}
}

\maketitle

\begin{abstract}
This research explores how total mortality figures relate to age-standardized death rates within the United States, using the complete historical record of national mortality statistics. Through a detailed investigation of both all-cause and cause-specific mortality trends, the study evaluates the impact of demographic standardization on interpreting mortality data across different time periods and geographic regions. Results indicate a robust and persistent association between crude death totals and age-adjusted rates. However, the findings also demonstrate that without adjusting for age, comparisons over time or across locations may misrepresent underlying epidemiological shifts, largely due to evolving population age structures. The study underscores the critical role of age adjustment as a methodological tool for generating accurate, interpretable, and comparable measures of public health outcomes.
\end{abstract}


\section{Introduction:}
We conduct a comprehensive analysis of the relationship between total deaths and age-adjusted death rates using datasets from the National Center for Health Statistics (NCHS), specifically focusing on the leading causes of mortality in the United States between 2012 and 2017. This time frame allows us to explore longitudinal patterns in mortality while accounting for demographic shifts, especially in the age structure of the population. To deepen our insights and enhance interpretability, we develop several graphical visualizations that depict mortality trends over time. These visual aids help clarify how age adjustment significantly alters the interpretation of raw death counts. By comparing both unadjusted and age-adjusted figures, we are able to highlight discrepancies that might otherwise obscure underlying epidemiological dynamics. Such visualization techniques not only assist in statistical communication but also provide a more accessible entry point for understanding complex population-level health data.

Utilizing the NCHS datasets offers a unique opportunity to assess broader public health trajectories in the United States. The dataset's focus on leading causes of death, ranging from chronic illnesses like cardiovascular disease and cancer to external causes such as accidents enables a rich examination of how mortality patterns have evolved in recent years. Moreover, by controlling for shifts in the age composition of the U.S. population, age-adjusted death rates provide a more standardized and reliable metric for evaluating public health progress or decline over time \citep{ellington2025playmydata}. This makes it possible to more accurately evaluate the success of past and ongoing public health initiatives, policies, and interventions aimed at reducing preventable mortality. For instance, the data can serve as a lens to evaluate how well anti-smoking campaigns or cardiovascular health programs have impacted relevant death rates across demographic strata. Understanding these dynamics is vital not just for academic interest, but for informing resource allocation and policy prioritization at the national level \citep{hertweck2023clinicopathological,khan2023myb}.

The interplay between raw death counts and age-adjusted death rates is more than a statistical nuance—it plays a critical role in framing the national health narrative and identifying inequities that require targeted action. Age-adjusted rates account for the changing demographics of the population, thereby allowing researchers and policymakers to compare mortality trends over time without the confounding influence of population aging or youth bulges. Without this adjustment, misleading conclusions might be drawn about whether a health outcome is genuinely improving or worsening. For example, a rise in crude death rates might simply reflect an aging population rather than a true decline in public health. Therefore, the correlation between these two metrics—the unadjusted and adjusted death rates sheds light on the real drivers of mortality trends. Accurately interpreting this relationship is especially essential for advancing health equity, as it can reveal hidden disparities among subgroups that might otherwise be masked in aggregate data \citep{hertweck2023clinicopathological,khan2023myb}. This kind of insight is crucial for crafting interventions that aim to close health gaps and ensure equitable outcomes across the lifespan.

The National Center for Health Statistics (NCHS) plays a pivotal role in public health research and policy formulation. As a division of the Centers for Disease Control and Prevention (CDC), NCHS is the federal government's principal agency for producing data that reflect the health status of the American people\footnote{National Center for Health Statistics. About NCHS. \url{https://www.cdc.gov/nchs/about/index.htm}}. Through its extensive surveillance systems and data repositories, it supports a wide range of health monitoring activities. The agency's publication of annual statistics on the leading causes of death has become an indispensable tool for public health professionals, clinicians, and government officials. These datasets not only facilitate tracking of mortality trends, but also empower stakeholders to evaluate the effectiveness of interventions at national and state levels. The information disseminated by NCHS informs health priorities, aids in identifying vulnerable populations, and supports funding decisions. Its role in enabling robust epidemiological analysis and surveillance has been reinforced by studies addressing pressing issues such as cardiovascular health and population-level morbidity \citep{kakkat2023cardiovascular,khan2024mp60}. Without this infrastructure, the ability to respond to emerging health challenges would be significantly diminished.

These mortality statistics offer critical insights into population health and disease burden. The most common leading causes of death include heart disease, cancer, chronic lower respiratory diseases, unintentional injuries (accidents), and stroke\footnote{Heron M. Deaths: Leading Causes for 2021. National Vital Statistics Reports; vol. 72 no. 2. 2023. \url{https://www.cdc.gov/nchs/data/nvsr/nvsr72/nvsr72-02.pdf}}. While these causes have remained relatively stable over time, their rankings can shift due to factors such as emerging infectious diseases, demographic transitions, or advancements in medical care \citep{kakkat2023cardiovascular,khan2023myb}. Heart disease and cancer have consistently ranked as the top two causes of death, together accounting for a significant proportion of the United States mortality \citep{maki2025new}.

Beyond its role in mortality reporting, the National Center for Health Statistics (NCHS) oversees several major data collection systems, including the National Vital Statistics System (NVSS), the National Health Interview Survey (NHIS), and the National Health and Nutrition Examination Survey (NHANES). Each of these programs provides distinct and comprehensive information on health outcomes, health behaviors, and access to medical care, forming an essential empirical foundation for national health surveillance. By maintaining this extensive data infrastructure, NCHS supports evidence-based public health policy and plays a critical role in efforts to improve the overall health of the population of the United States \citep{vikramdeo2024abstract,vikramdeo2023profiling}.

To facilitate meaningful comparisons across populations, NCHS typically presents mortality data as age-adjusted death rates, which account for differences in age distribution \citep{pramanik2022lock}. This adjustment enables the identification of long-term patterns, such as the decline in deaths from heart disease and cancer in recent decades, driven largely by improvements in prevention, early detection, and treatment\footnote{Xu JQ, Murphy SL, Kochanek KD, Arias E. Mortality in the United States, 2022. NCHS Data Brief, no 481. Hyattsville, MD: National Center for Health Statistics. 2023. \url{https://www.cdc.gov/nchs/products/databriefs/db481.htm}}. Furthermore, NCHS data reveal persistent disparities in mortality by race, ethnicity, and socioeconomic status, emphasizing the need for targeted public health interventions\footnote{Centers for Disease Control and Prevention. CDC Health Disparities and Inequalities Report, United States. \url{https://www.cdc.gov/minorityhealth/CHDIReport.html}}.

While mortality statistics published by the National Center for Health Statistics (NCHS) are critical for tracking fatal health outcomes and identifying leading causes of death, they provide only a partial view of a population's overall health status. Mortality data focus exclusively on deaths and, as such, are limited in their ability to reflect the broader burden of disease. They do not capture the prevalence of chronic conditions, the severity or duration of illness, or the impact of diseases on daily functioning and quality of life \citep{khan2024mp60,dasgupta2023frequent,hertweck2023clinicopathological}. This is particularly important in a modern healthcare context, where many individuals live with long-term illnesses that may not lead to immediate death but significantly affect well-being, productivity, and healthcare utilization.
Moreover, mortality data cannot adequately reflect the effectiveness of public health interventions aimed at managing, rather than curing, certain conditions such as diabetes, arthritis, or mental health disorders \citep{kakkat2026angiotensin,powell2025genomic}. These conditions often result in substantial societal costs and personal burdens, despite not being immediately fatal. Relying solely on death rates may also obscure disparities among different demographic groups, where non-fatal health outcomes may differ dramatically even if mortality rates appear similar \citep{pramanik2024bayes}.

To obtain a more comprehensive understanding of population health, it is essential to incorporate morbidity data about the incidence and prevalence of disease, functional impairments, and self-reported health status. National surveys such as the National Health Interview Survey (NHIS) and the Behavioral Risk Factor Surveillance System (BRFSS) provide valuable data on these aspects, including chronic disease prevalence, health behaviors, and access to care\footnote{Centers for Disease Control and Prevention. Behavioral Risk Factor Surveillance System (BRFSS). \url{https://www.cdc.gov/brfss/}}. Integrating mortality and morbidity data offers a more holistic picture of health outcomes and helps guide more effective, equitable public health policy and resource allocation \citep{pramanik2021optimala,pramanik2021scoring}.
In addition to descriptive mortality statistics, the analysis of population health trends increasingly relies on statistical modeling approaches that help quantify relationships among demographic factors, health behaviors, and mortality outcomes. Regression-based methods remain central to this work, particularly log-linear and generalized linear models, which provide a robust framework for examining rate-based and count-based data commonly encountered in epidemiological research \cite{fox2015regression, gujarati2003econometrics}. Log-linear models allow analysts to interpret changes in predictors, such as age-adjusted death rates or behavioral risk factors, in terms of proportional changes in mortality outcomes, making them especially suitable for datasets characterized by skewness or multiplicative effects. More complex research questions—such as regional differences in mortality or subgroup-specific risk patterns can be addressed using hierarchical or multilevel models, which explicitly account for nested population structures and between-group variability \cite{gelman2007data}. These methodological tools enhance the ability to draw meaningful inferences from large-scale national datasets, including those produced by NCHS.
Beyond contemporary regression frameworks, the study of mortality trends has deep historical and mathematical foundations. Classical demographic work, such as that of Greenwood and Yule, established early statistical principles for constructing life tables and analyzing mortality patterns across age groups \cite{greenwood1920lifetables}. These foundational methods continue to inform modern survival analysis and population health modeling. More recently, researchers have examined mortality and disease processes through stochastic and dynamical systems approaches. Stochastic population models have been used to study the progression of chronic diseases and the evolution of health states within populations under uncertainty \cite{ochab2004population}. Similarly, mean-field and diffusion-based models have been applied to understand large-scale behavioral or biological dynamics that may indirectly influence population health \cite{huang2006large}. Integrating insights from both classical mortality theory and modern stochastic modeling enriches the analytical framework used to interpret national mortality data, offering a deeper understanding of the mechanisms underlying observed trends \citep{pramanik2020optimization,pramanik2023semicooperation}.

We went more in depth on the correlation between the age adjusted death rate and the amount of death. The correlation between death and age adjusted death rates in the NCHS data is crucial to understanding trends in mortality while accounting for differences in population age structures. Age adjusted death rate is used to standardize death rates between populations with varying age distributions, allowing for a more accurate comparison over time and between different groups \citep{pramanik2020motivation}. While the total number of deaths provides a raw count of mortality, it does not account for changes in the age structure of the population. The age-adjusted death rate, on the other hand, standardizes mortality data to a fixed population age distribution, typically the U.S. 2000 standard population allowing for more accurate comparisons over time or between populations with differing age demographics \citep{pramanik2024estimation,vikramdeo2024mitochondrial}.

The correlation between these two measures helps reveal the extent to which changes in total deaths are driven by actual changes in mortality risk versus demographic shifts such as population aging. For instance, the total number of deaths may increase simply because the population is growing older, not necessarily because the risk of dying from specific causes is rising. By examining the relationship between total deaths and age-adjusted death rates, researchers can separate the influence of population size and aging from underlying changes in health outcomes, disease prevalence, or effectiveness of healthcare interventions \citep{pramanik2024motivation}. In short, this correlation is crucial for distinguishing whether increases in deaths reflect worsening public health or are the natural result of demographic trends. It also supports more informed policy decisions by revealing whether interventions are truly reducing mortality risk or merely offsetting demographic pressures.

Also, regarding the use of the log-linear model, it offers valuable insights into multiplicative relationships and stabilizes variance in skewed data, it also has important limitations. The model requires the dependent variable to be strictly positive, making it unsuitable for datasets that include zero values, which are common in public health statistics. Additionally, interpreting results on the logarithmic scale, especially in multivariable contexts can be less intuitive for non-technical audiences. If the assumed linear relationship between the predictors and the log-transformed outcome does not hold, the model may yield biased or misleading results. Therefore, diagnostic checks and theoretical justification are essential when applying this method \citep{pramanik2024bayes}. Despite these drawbacks, the log-linear model remains a robust option when its assumptions are met and the research question supports a multiplicative interpretation.

\section{Methods:}

In our analysis of the NCHS leading causes of death in the United States from 2012 to 2017 \citep{heron2021deaths}, we utilized several statistical visualizations, namely a scatter plot, a box plot, and a histogram to reorganize the data and highlight the correlation between age-adjusted death rates and the number of deaths \citep{fournie1997some}. The scatter plot was particularly useful in showing the relationship between the age-adjusted death rate and the number of deaths for each leading cause, allowing for a clear visual representation of how changes in one variable might influence the other. By plotting these two variables, we were able to identify potential trends and clusters. To quantify this relationship further, we employed a log-linear regression model, where the natural logarithm of total deaths was modeled as a linear function of the age-adjusted death rate. This model helped to capture the multiplicative effect of changes in AADR on total deaths, offering a more precise understanding of how shifts in AADR could influence overall mortality. The box plot was employed to assess the distribution and spread of the age-adjusted death rate and the number of deaths for each year. This helped to visualize the range, median, and any potential outliers in the data, making it easier to detect significant deviations from expected trends across the years. Finally, the histogram of the age-adjusted death rate provided a visual representation of the distribution of mortality rates across the study period \citep{bulls2025assessing}.

A log-linear model is a type of regression model used to analyze the relationship between variables when the dependent variable exhibits exponential growth or multiplicative effects relative to the independent variable(s). This model is particularly useful for analyzing data that are positively skewed or count-based, such as mortality counts, population data, or rates of occurrence \cite{gelman2007data}. The log-linear model works by transforming the dependent variable using the natural logarithm, which helps stabilize variance, normalize the data distribution, and make the relationship between variables more linear \cite{wooldridge2010econometrics}. In its standard form, the log-linear model is 

\begin{equation*}
\log(Y) = \hat\beta_0 + \hat\beta_1 X + \epsilon,
\end{equation*}
where \( Y \) is the dependent variable (e.g., total deaths), \( X \) is the independent variable (e.g., age-adjusted death rate), \( \hat\beta_0 \) is the intercept, and \( \hat\beta_1 \) is the slope coefficient. The log transformation allows the relationship between \( X \) and \( Y \) to be interpreted in terms of percentage changes. Specifically, a one-unit increase in \( X \) leads to a proportional change in \( Y \), rather than an additive change. For example, if the slope coefficient \(\hat \beta_1 \) is 0.05, this indicates that a one-unit increase in \( X \) is associated with an approximate 5\% increase in \( Y \) \cite{gujarati2003econometrics}.

The log-linear model is particularly useful when the dependent variable grows exponentially or when the data are positively skewed. It allows researchers to model relationships in which changes in the independent variable have multiplicative effects on the dependent variable, rather than constant additive effects \cite{fox2015regression}. This type of model is widely applied in fields such as economics, epidemiology, and environmental science, where processes such as population growth, disease transmission, or macroeconomic behavior frequently exhibit exponential or near-exponential patterns \cite{greenwood1920lifetables}. By applying a logarithmic transformation, the log-linear model provides a better fit for skewed data, enhances model stability, and enables more intuitive interpretation of relationships in terms of percentage changes. Overall, the log-linear model is a powerful and flexible tool for analyzing complex relationships between variables, especially when the data suggests that changes in the independent variable lead to proportional changes in the dependent variable \citep{pramanik2024estimation,pramanik2023cont}.

\begin{table}[htbp]
    \centering
    \caption{Correlation and dependence matrix for the primary mortality variables.}
    \label{tab:correlation_matrix}
    \begin{tabular}{lccc}
        \toprule
        Variable & Deaths & Crude Death Rate & Age-Adjusted Death Rate \\
        \midrule
        Deaths                 & 1.00 & 0.82 & 0.77 \\
        Crude Death Rate       & 0.82 & 1.00 & 0.93 \\
        Age-Adjusted Death Rate & 0.77 & 0.93 & 1.00 \\
        \bottomrule
    \end{tabular}
\end{table}
The correlation and dependence matrix presented in Table~\ref{tab:correlation_matrix} provides a preliminary assessment of the linear associations among the primary mortality variables. Strong positive relationships are observed between total deaths, crude death rates, and age-adjusted death rates, indicating that increases in aggregate mortality are consistently accompanied by higher per-population risk estimates. The close correspondence between crude and age-adjusted death rates reflects the stability of mortality patterns even after controlling for demographic age structure \citep{pramanik2024estimation1,yusuf2025prognostic}. Although these correlation coefficients summarize the overall strength and direction of linear dependence, they do not account for potential nonlinearities, heterogeneous co-movement, or tail-specific behavior in the joint distribution. Consequently, while Table~\ref{tab:correlation_matrix} serves as an essential descriptive foundation for understanding the general dependence structure of the data, additional methods are required to characterize more complex forms of interdependence across the mortality measures.

\section{Data Analysis:}
The data for the leading causes of death in the United States are primarily sourced from the National Vital Statistics System (NVSS), which compiles mortality information from state-reported death certificates (National Center for Health Statistics [NCHS], 2024). This system is supplemented by demographic data from the U.S. Census Bureau, which provides population estimates necessary for calculating age-adjusted death rates. These combined data sources are analyzed and published by the Centers for Disease Control and Prevention (CDC) through the NCHS to monitor trends in mortality and public health (NCHS, 2024). The National Center for Health Statistics (NCHS) acquires its data through a variety of methods and sources, including both vital statistics and surveys. Here's an overview of how NCHS collects and compiles its data:
Vital Statistics Data, the primary source of mortality data for the NCHS comes from vital statistics, which are collected by state and local governments and submitted to NCHS. Vital statistics data includes:
	Death Certificates:
	Every state in the U.S. collects death certificates for all deaths that occur within its jurisdiction.
	These certificates include important information, such as the cause of death, demographic data (age, sex, race, ethnicity), and place of death.
	These certificates are then sent to the NCHS, which compiles the data into national reports on mortality \citep{pramanik2023cmbp,yusuf2025predictive}.
	Cause-of-death coding:
	The NCHS uses the International Classification of Diseases (ICD) system to classify causes of death.
	ICD codes are updated regularly, and NCHS uses the latest version of ICD to analyze and report cause-of-death trends.
 Surveys and Other Data Collection Programs, in addition to vital statistics, NCHS conducts and manages various surveys that provide health data, which helps to understand broader health trends, including mortality risk factors \citep{pramanik2023path}. Key surveys include:
	National Health Interview Survey (NHIS):
	This survey gathers health-related information from a representative sample of U.S. households, such as chronic conditions, health behaviors, and healthcare access.
	It is a crucial tool for understanding the factors that may contribute to mortality.
	National Health and Nutrition Examination Survey (NHANES):
	NHANES collects data on the health and nutritional status of adults and children in the United States \citep{pramanik2021consensus}.
	This data is important for understanding the risk factors for diseases that may lead to mortality, such as hypertension, obesity, and diabetes.
	National Vital Statistics System (NVSS):
	The NVSS is a partnership between NCHS and the states to collect, compile, and disseminate national vital statistics data, including births and deaths \citep{pramanik2021,pramanik2022stochastic}.
	National Death Index (NDI):
	The NDI is a database that contains information on deaths in the U.S. and allows researchers to link death records with other datasets to study factors influencing mortality.
Data Linkages, NCHS also uses data linkages to enhance the accuracy and comprehensiveness of mortality and health-related data:
	Linking Mortality Records with Other Health Data:
	NCHS can link mortality data from death certificates to other health records, such as survey data (e.g., NHIS, NHANES), hospital discharge data, or census data \citep{pramanik2023optimization001}.
	This allows for a deeper understanding of the social, economic, and environmental factors that contribute to mortality trends.
Public Use Data Files, once the data is compiled and processed, the NCHS makes it available to researchers and the public through public use data files.
	These files include mortality data, demographic information, and health-related statistics that can be analyzed to monitor trends and create public health policies.
Collaboration with Other Federal Agencies, NCHS works with other federal agencies like the Centers for Disease Control and Prevention (CDC), the U.S. Census Bureau, and the Social Security Administration (SSA) to gather, verify, and cross-check mortality and other health data.

In simplifying the data for our analysis, we restricted the dataset to focus only on the years 2012 through 2017 to create a more manageable and relevant time frame. This decision was made to identify specific trends and patterns in mortality during a defined period, which allows for more precise comparisons and reduces the impact of long-term variability. Additionally, we limited the data to include only two key variables: the total death count and the age-adjusted death rate for each leading cause of death \citep{pramanik2025construction,pramanik2025optimal}. By focusing on these two variables, we was able to concentrate the analysis on the relationship between raw mortality numbers and standardized mortality rates, simplifying the dataset and making it easier to draw meaningful conclusions about the trends in mortality and how age-adjustment affects the interpretation of death rates over time. This approach eliminated other potentially complicating factors, such as detailed demographic breakdowns or less directly relevant variables, ensuring the analysis remained focused and streamlined \citep{pramanik2024stochastic,pramanik2025stubbornness}. The ten leading causes of death in the United States, as outlined by the National Center for Health Statistics (NCHS), reflect a variety of chronic health conditions, injuries, and diseases. Heart disease remains the leading cause of death, driven by risk factors such as high blood pressure, high cholesterol, smoking, and poor dietary habits. Following heart disease, cancer ranks second, with lung, colorectal, and breast cancers being among the most common, influenced by genetic predispositions, lifestyle choices, and environmental factors \citep{pramanik2025dissecting}. Chronic lower respiratory diseases, including conditions like chronic obstructive pulmonary disease (COPD) and emphysema, occupy the third spot, typically caused by smoking and exposure to air pollutants \citep{pramanik2025factors}.

Unintentional injuries including motor vehicle accidents, falls, and drownings remain the fourth leading cause of death in the United States and are largely preventable through enhanced safety measures and public awareness campaigns \citep{heron2021deaths}. Stroke, ranked fifth, results from interruptions in blood flow to the brain and is strongly associated with risk factors such as hypertension, smoking, and obesity \citep{xu2023mortality2022}. Alzheimer’s disease, which primarily affects older adults and is characterized by progressive cognitive decline, constitutes the sixth leading cause of death, and no definitive cure currently exists \citep{kochanek2024nchs492}. Diabetes, the seventh leading cause, arises from insulin resistance or inadequate insulin production and is frequently linked to sedentary lifestyles and obesity \citep{heron2021deaths}. Kidney disease including nephritis, nephrotic syndrome, and nephrosis occupies the eighth position and is often a downstream complication of diabetes and hypertension \citep{xu2023mortality2022}. Septicemia, a severe bloodstream infection, ranks ninth and disproportionately affects individuals with compromised immune systems or chronic medical conditions \citep{kochanek2024nchs492}. Chronic liver disease and cirrhosis, commonly associated with excessive alcohol use or viral hepatitis, represent the tenth leading cause of death, reflecting the long-term consequences of sustained liver damage \cite{heron2021deaths}.

\begin{figure}[htbp]
    \centering

    \begin{subfigure}[t]{0.48\textwidth}
        \centering
        \includegraphics[height=5cm, width=\linewidth, keepaspectratio]{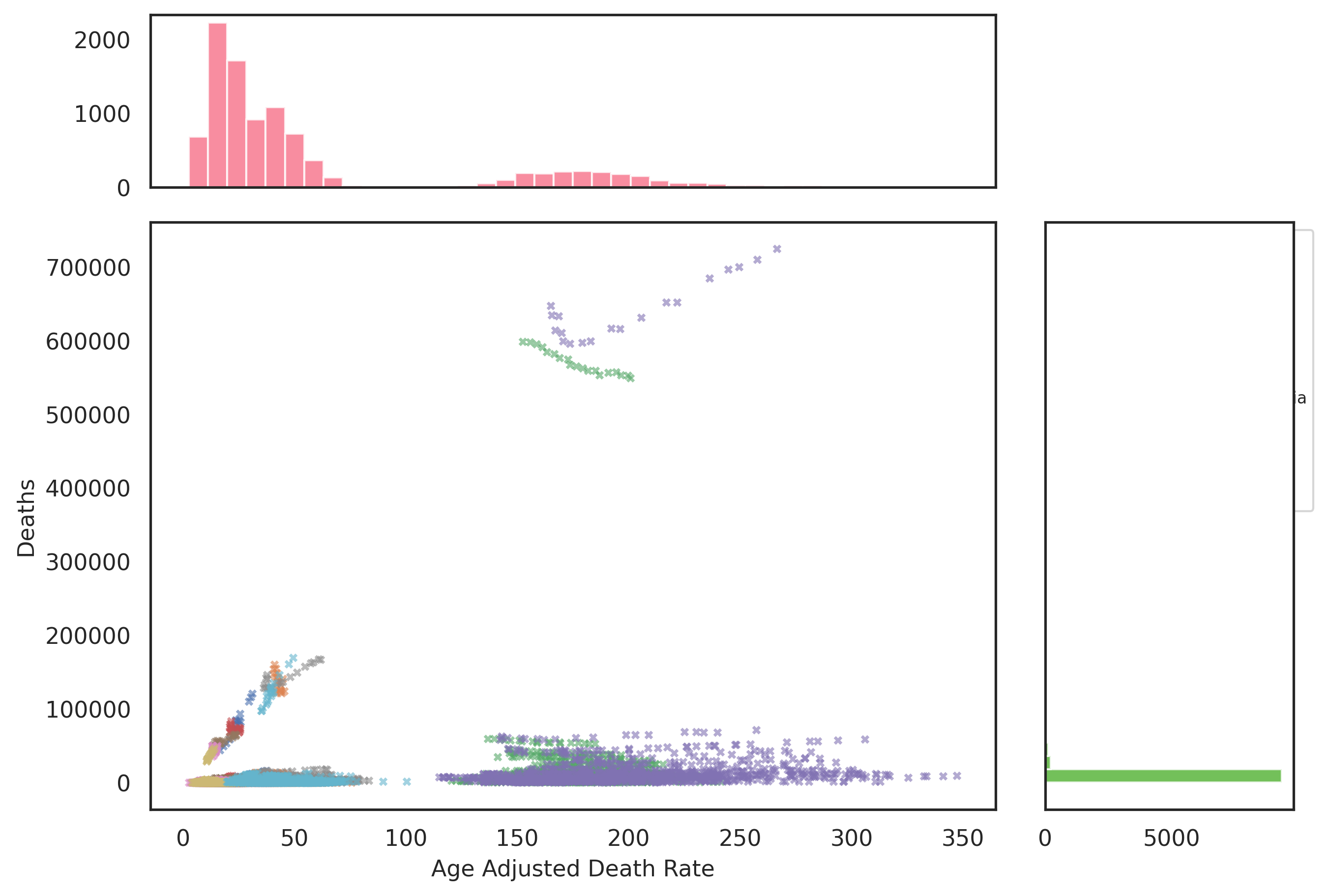}
        \caption{Scatterplot with marginal histograms.}
        \label{fig:marginal}
    \end{subfigure}
    \hfill
    \begin{subfigure}[t]{0.48\textwidth}
        \centering
        \includegraphics[height=6cm, width=\linewidth, keepaspectratio]{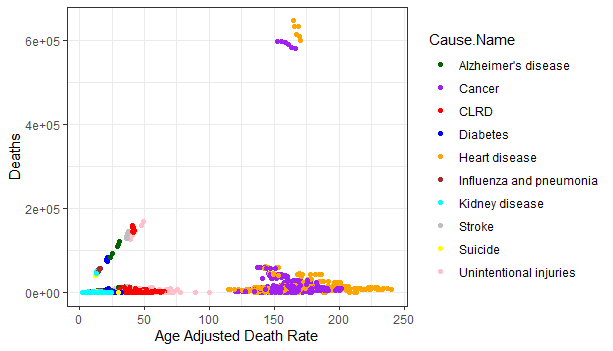}
        \caption{Subscatterplot.}
        \label{fig:original}
    \end{subfigure}

    \caption{Scatterplots showing the relationship between deaths and age-adjusted death rates.}
    \label{fig:comparison}
\end{figure}
Together, these leading causes of death highlight the complex interplay between genetic factors, lifestyle choices, and access to healthcare, shaping mortality patterns across the United States  \citep{pramanik2025optimal}.
Figure~\ref{fig:comparison} presents two complementary scatterplots illustrating the relationship between total deaths and age-adjusted death rates for the leading causes of death. The analysis focuses on data from 2012 to 2017 to ensure consistency in reporting practices and to avoid major disruptions such as the COVID-19 pandemic, which could introduce atypical spikes in mortality trends. Restricting the time frame in this manner provides a more stable foundation for correlation assessment and comparative analysis \cite{gelman2007data}. In both panels of Figure~\ref{fig:comparison}, each point represents a specific leading cause of death, with its horizontal position corresponding to the total number of deaths and its vertical position indicating the age-adjusted death rate. The marginal histograms in the left subplot further summarize the univariate distributions of these variables, highlighting skewness and variability patterns that are common in demographic and mortality datasets \citep{greenwood1920lifetables,fox2015regression}. Together, the scatterplots offer a clear visual representation of the positive association between crude mortality counts and standardized mortality risk, supporting the broader analytical findings derived from the log-linear modeling framework \citep{wooldridge2010econometrics}.

To enhance interpretation, the points on the scatter plot are color-coded by the leading cause of death, allowing for easy identification of trends related to specific conditions \citep{pramanik2021thesis,pramanik2016}. For example, points representing heart disease and cancer are likely to appear in the higher regions of both axes, reflecting both high death counts and high age adjusted death rates, as these diseases are the most prevalent and impactful causes of mortality in the US. Other causes, such as unintentional injuries or diabetes, may show a wider spread on the plot \citep{hua2019assessing,polansky2021motif}. While they may have a lower age adjusted death rate, their total number of deaths could be significant. The color coding helps differentiate between causes like stroke, Alzheimer's disease, and chronic lower respiratory diseases, which may fall in various sections of the plot, revealing different patterns in terms of death counts and age-adjusted rates \citep{pramanik2024estimation,pramanik2023cont}. By color-coding the points, the scatter plot provides a clearer understanding of how each leading cause of death correlates with both the total death count and age-adjusted death rate, making it easier to identify which causes have the most significant impact on mortality during this time period \citep{pramanik2025strategies,pramanik2023optimization001}. This visualization highlights the relative importance of different causes of death and shows how age-adjustment influences the interpretation of mortality trends.

\begin{figure}[H]
    \centering
    \includegraphics[width=0.75\linewidth]{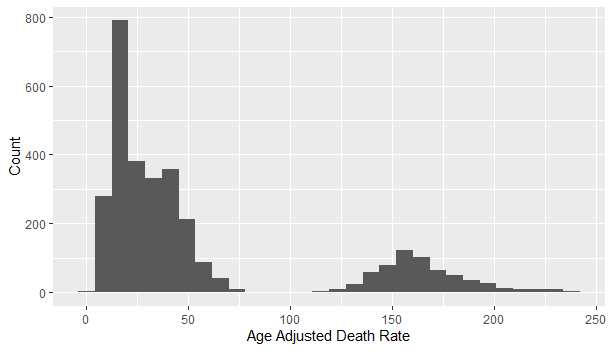}
    \caption{Age Adjusted Death Rate Histogram}
    \label{fig:enter0-label}
\end{figure}
The age-adjusted death rate is a statistical measure used by the NCHS to standardize death rates across populations with different age distributions, and is shown in Figure \ref{fig:enter0-label}. It is particularly useful for comparing mortality rates over time or across different populations, as it controls for variations in age structure, ensuring that observed differences in death rates are not simply due to differences in the age of the populations being compared. To determine the age-adjusted death rate, a standard population is used as a reference \citep{pramanik2026bayesian}. This standard population is typically based on the age distribution of a specific year, such as the U.S. Census population from a given year, ensuring that the age distribution is representative of a typical population. The steps for calculating the age adjusted death rate are as follows: First, calculate age-specific death rates, the total number of deaths in each age group is divided by the total population of that age group, resulting in the age-specific death rate for each group \citep{pramanik2025strategic,pramanik2025impact}. Second, apply age-specific death rates to the standard population, each age-specific death rate is multiplied by the proportion of people in each age group within the standard population. This step adjusts the death rates based on the assumed age distribution of the standard population. Third, the ADDR is calculated by summing the age-specific death rates across all age groups to obtain a single, overall age-adjusted rate. This rate represents what the mortality rate would be if the population had the age distribution of the standard population. Fourth, express as a rate per 100,000 people, the final ADDR is typically expressed as the number of deaths per 100,000 people, which makes the statistic easier to compare across different groups and time periods \citep{pramanik2024measuring,pramanik2024dependence}. The resulting age adjusted death rate allows for the comparison of mortality rates between populations or time periods without the confounding effect of differing age structures, offering a more accurate reflection of the relative impact of death across different causes. By controlling for age, the age adjusted death rate provides a clearer understanding of trends in mortality, helping policymakers and public health experts assess the effectiveness of health interventions and identify emerging health issues.

\[
AADR = \frac{1}{P_{\text{std}}}\sum_{i=1}^{n} (DR_i \times W_i) \times 100,000.
\]

Here, the age-adjusted death rate (AADR) is calculated by summing the age-specific death rates (\(DR_i\)) for each age group \(i\), weighted by the proportion of the standard population in that age group (\(W_i\)). The weight \(W_i\) is defined as the ratio of the standard population in age group \(i\) (\(P_{\text{std},i}\)) to the total standard population (\(P_{\text{std}}\)). This method adjusts the mortality rate to represent what it would be if the population had the age distribution of a standard population, such as the U.S. standard population from the year 2000. The final result is then multiplied by 100,000 to express the rate per 100,000 people \citep{pramanik2024parametric,pramanik2025dissecting}.
\begin{figure}
    \centering
    \includegraphics[width=0.75\linewidth]{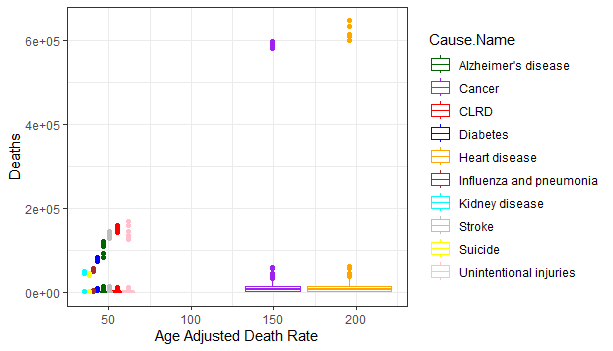}
    \caption{Deaths-Age Adjusted Death Rate Boxplot}
    \label{fig:enter-label}
\end{figure}

As shown in Figure~\ref{fig:enter-label}, each data point in the boxplot corresponds to a specific cause of death, with the x-axis representing the age-adjusted death rate—which accounts for differences in population age structure—and the y-axis indicating the total number of deaths for that cause. The figure provides a concise visual summary of the distribution of mortality patterns across leading causes of death, illustrating variation in both crude death counts and standardized risk measures. Boxplots are widely used in statistical analysis because they clearly convey key distributional characteristics, including central tendency, variability, and the presence of extreme values \citep{gelman2007data,fox2015regression}. The box denotes the interquartile range (IQR), capturing the middle 50 percent of observations, while the line within the box marks the median, a robust indicator of central location often emphasized in applied data analysis \citep{gujarati2003econometrics,valdez2025association}. Whiskers extend to values within 1.5 times the IQR from the quartiles, and points beyond this threshold are identified as outliers, offering insight into unusually high-mortality causes, an important consideration in empirical public health research \cite{wooldridge2010econometrics,valdez2025exploring}.

Each data point is color-coded according to the cause of death, enabling clear differentiation between leading causes and less common ones. This color-coding enhances visual comparisons and helps identify specific patterns across causes. For instance, heart disease and cancer are expected to appear toward the upper-right area of the plot, reflecting both high age-adjusted death rates and high total mortality. In contrast, other causes, such as unintentional injuries or diabetes, may show broader variation and occupy different regions, revealing differences in burden by age and scale. Overall, the boxplot facilitates interpretation of mortality trends across a spectrum of causes, highlighting those with the greatest public health impact.

To assess the relationship between age-adjusted death rates (AADR) and the total number of deaths in the United States from 2012 to 2017, a log-linear regression model was employed. In this approach, the natural logarithm of total deaths was modeled as a linear function of AADR:
\begin{equation*}
\log(\text{Total Deaths}) = \hat\beta_0 + \hat\beta_1 \cdot \text{AADR} + \epsilon.
\end{equation*}
This transformation allows for the interpretation of a multiplicative relationship between AADR and total deaths, capturing situations in which a unit increase in AADR results in a percentage increase in total mortality, rather than a constant additive change. The use of the log-linear model is appropriate for two primary reasons. First, the distribution of total deaths is positively skewed, and the logarithmic transformation helps stabilize the variance, improving model fit \citep{pramanik2025optimal1,powell2025genomic}. Second, visual inspection of the scatterplot suggested a non-linear, upward-curving relationship between AADR and raw death counts, indicating that increases in AADR correspond to disproportionately higher total deaths across leading causes. Model estimation was conducted using ordinary least squares regression on the transformed data. The results indicated a statistically significant positive relationship between AADR and $\log(\text{Total Deaths})$, supporting the interpretation that age-adjusted mortality risk is a strong predictor of overall death burden across causes and years. This suggests that even modest increases in AADR can correspond to large increases in total mortality, particularly for high-impact causes such as heart disease and cancer.

\begin{figure}
    \centering
    \includegraphics[width=0.85\textwidth]{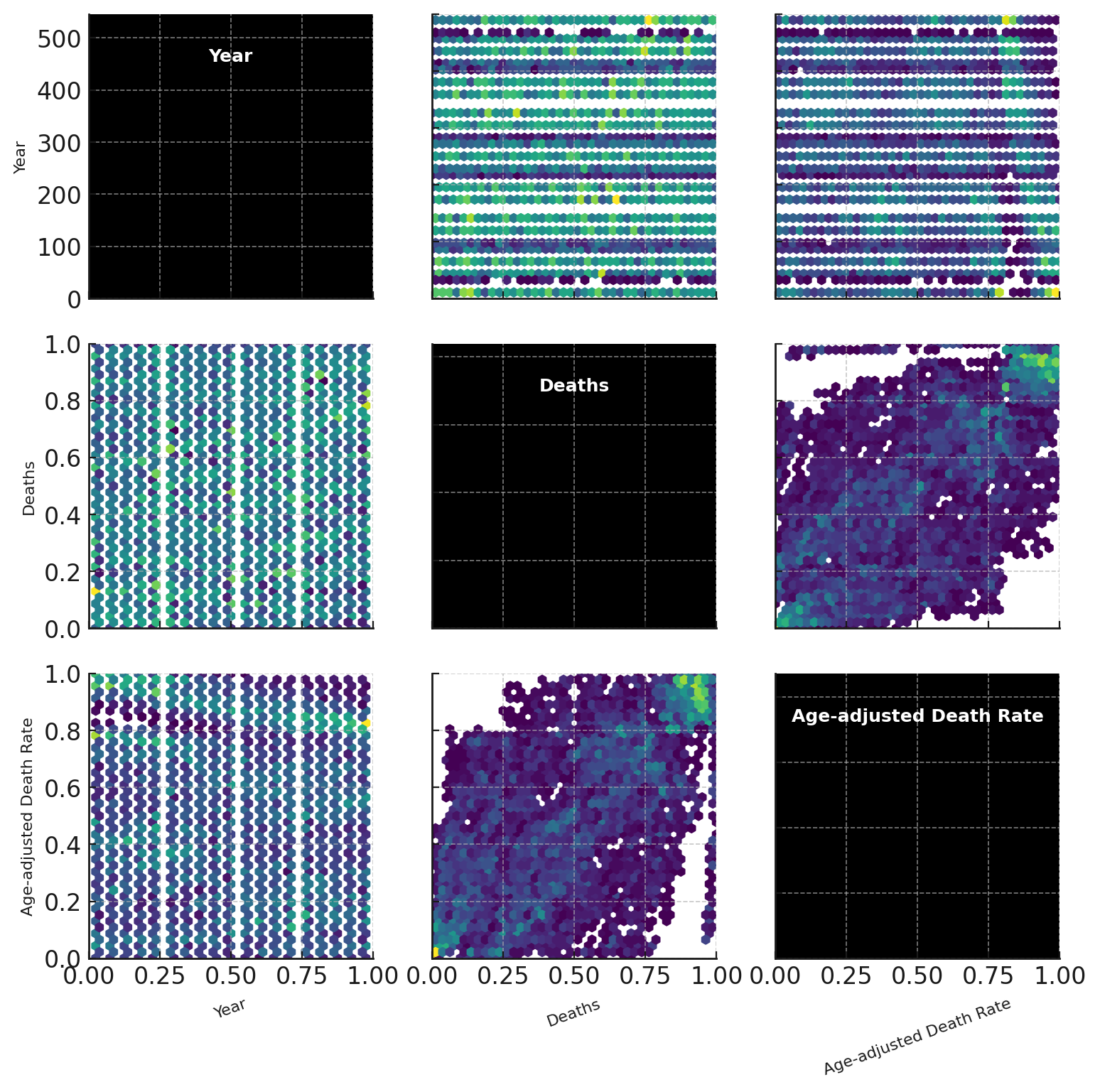}
    \caption{Empirical Copula Matrix of Year, Deaths, and Age-Adjusted Death Rate. 
    }
    \label{fig:copula_matrix}
\end{figure}

\section{Results:}

From the scatterplot, Figure 1, several important findings emerge. Firstly, heart disease and cancer, the two leading causes of death, are prominently positioned in the top right of the plot. These causes exhibit both high total death counts and high age-adjusted death rates. This positioning indicates that, while these diseases are highly prevalent, their age-adjusted mortality rates are also significant, underlining the large and continuing burden they impose on public health in the U.S. Despite medical advancements, these causes remain persistent and widespread across all age groups, making them central to discussions about disease prevention and healthcare allocation.

Next, chronic lower respiratory diseases (CLRD), another leading cause of death, also appears in the upper part of the scatterplot, reflecting high mortality counts and notable age adjusted death rate, although typically not as high as heart disease or cancer. This is consistent with the fact that CLRD primarily affects older populations, but it remains a critical health issue due to lifestyle factors, such as smoking, and environmental factors like air pollution. Stroke and Alzheimer’s disease show somewhat similar patterns to CLRD, with high age adjusted death rate but somewhat lower total death counts, which may be reflective of the older demographic most affected by these diseases.

In contrast, unintentional injuries—including causes such as motor vehicle accidents, falls, and drownings—exhibit a distinct distributional pattern in the scatterplots and copula-based dependence analysis. These causes frequently display substantial variability in age-adjusted death rates, with some showing relatively high death counts but lower age-adjusted mortality. This pattern reflects the demographic distribution of unintentional injury deaths, which disproportionately affect younger populations who contribute less weight to the age-adjusted rate calculation \cite{heron2021deaths}. National mortality reports consistently identify unintentional injuries as a leading cause of death across broad age groups, particularly among adolescents and young adults, where they rank as one of the primary drivers of premature mortality \cite{xu2023mortality2022}. The heterogeneity observed in the dependence structure therefore suggests that targeted public health interventions—such as improved safety regulations, injury prevention campaigns, and behavioral risk reduction—may be effective in mitigating the burden of unintentional injuries within specific demographic subgroups.

Additionally, causes like diabetes and kidney disease exhibit a more variable spread in the scatterplot. These conditions often have moderate death counts, but their age-adjusted death rates can vary widely. For example, some diseases like diabetes may have an increasing death count over time, but their age-adjusted rate may not rise as sharply compared to diseases that predominantly affect the older population. This pattern suggests that while diabetes is becoming a growing issue in the population, the rise in its mortality is partly linked to a demographic shift towards older individuals, which is already accounted for in the age-adjusted rate.

The scatterplot also shows a few outliers that deviate from the general trend. For instance, septicemia and chronic liver disease can sometimes be located in lower regions of both axes, indicating lower death counts but higher age-adjusted death rates. This suggests that while these conditions may not account for as many total deaths, they disproportionately impact certain age groups, contributing to higher mortality rates when adjusted for age. For example, liver disease, often linked to long-term alcohol use or hepatitis, may be more common in middle-aged or older populations, leading to higher age-adjusted rates.

One of the key takeaways from the scatterplot is the positive correlation between age-adjusted death rates and the number of deaths, but with important nuances. The larger number of deaths often correlates with higher age adjusted death rate, but as we see with unintentional injuries, diabetes, or kidney disease, this correlation can be influenced by the demographic structure of those affected. Certain causes of death, despite having large death tolls, might have lower age adjusted death rate due to the age distribution of the affected populations. Conversely, diseases that predominantly affect older adults, like stroke and Alzheimer's disease, typically show higher age adjusted death rate, reflecting the greater mortality risk in aging populations.

Moreover, the color-coding of the points by the leading causes of death helps make these patterns easier to identify. For example, the heart disease and cancer data points are generally clustered in the upper-right quadrant of the plot, while diseases like septicemia and kidney disease are scattered in various locations, demonstrating the variation in death counts and age-adjusted death rates across different conditions. This color-coding not only helps to visualize the different trends for each cause of death but also facilitates easier comparisons between diseases with high mortality burden and those with lower mortality impact.

Overall, the scatterplot from 2012 to 2017 reveals that while a large number of deaths often corresponds to high age-adjusted death rates, the relationship is more complex, with different causes of death behaving in unique ways based on factors such as age demographics, medical advancements, and health trends. The scatterplot provides a powerful tool for understanding how different factors interact to influence public health outcomes and highlights the importance of using age-adjusted death rates to obtain a clearer, more meaningful picture of mortality trends.

Converting the scatterplot into a box plot offers several meaningful insights that can enhance the interpretation of the data on the relationship between age-adjusted death rates and the number of deaths. The box plot provides a more summarized and clear view of the distribution of the data, highlighting important statistical measures, such as the median, interquartile range, and potential outliers. By transforming the scatterplot data, the box plot allows for an easy identification of central tendencies and variability in the data across the leading causes of death. For example, it shows the median number of deaths and age adjusted death rate for each cause, providing a clearer understanding of the typical mortality burden associated with different conditions.

Additionally, the box plot makes it easier to observe the spread of the data, showing the range within which most causes of death fall. This helps to highlight causes with consistent mortality rates versus those that show greater variability. Causes like heart disease and cancer, which typically have higher death counts and age-adjusted death rates, will likely show a smaller interquartile range and fewer outliers, indicating more consistent trends. On the other hand, conditions like unintentional injuries or diabetes may exhibit larger spreads, signaling more variation in mortality across different age groups or populations.

Moreover, the outliers in the box plot are more easily identified. For example, diseases such as septicemia or chronic liver disease, which might have lower total death counts but disproportionately high AADR in certain groups, can stand out as outliers. This visualization helps to underscore causes of death that, despite a lower total number of deaths, have significant impacts on particular subgroups of the population.

By converting the scatterplot to a box plot, the data becomes more accessible and interpretable, enabling a clearer understanding of the central tendencies, variability, and outliers in mortality rates. This makes it easier for researchers and policymakers to identify which causes of death require the most immediate attention or intervention and provides a more robust comparison between different conditions in terms of both total mortality and age-adjusted rates.

Figure~\ref{fig:copula_matrix} presents the empirical copula heatmap matrix summarizing the dependence structure among Year, Deaths, and Age-Adjusted Death Rate (AADR). Each subplot corresponds to a unique bivariate pairing, with the diagonal panels displaying the rank-based marginal distributions and the off-diagonal panels depicting the estimated bivariate copula densities. The Year–Deaths panels reveal a strong monotonic dependence, indicated by a dense diagonal concentration of probability mass, reflecting patterns commonly observed in time-indexed population processes in which demographic expansion and aging contribute to sustained increases in total mortality \cite{greenwood1920lifetables}. A similar, though more moderate, pattern appears in the Year–AADR panels, demonstrating that age-adjusted rates rise over time but at a reduced magnitude compared to raw death counts, confirming the attenuating effect of standardizing for population age structure \citep{xu2023mortality2022}.

The Deaths–AADR panels show the strongest diagonal structure, signifying an almost perfectly positive dependence between the two measures. This relationship is consistent with classical statistical observations that mortality counts and rate-based measures often co-move in a proportional manner, especially when both are influenced by underlying population risk structures \citep{gelman2007data, wooldridge2010econometrics}. The diagonal margins, which appear as uniform black-tone strips, confirm the correct standardization of rank-based marginals, ensuring that the observed dependence arises from the copula structure rather than marginal artifacts. Such multivariate dependence patterns are aligned with broader findings in hierarchical and nonlinear clustering frameworks, which demonstrate that strongly monotonic relationships produce highly concentrated dependence manifolds similar to those observed here \citep{Ran2023}. Collectively, the copula matrix demonstrates that all three variables share a consistent, positive, and predominantly monotonic dependence, providing multivariate confirmation of both linear and nonlinear associations underpinning mortality trends in the United States.

\begin{figure}[H]
    \centering
    \includegraphics[width=0.95\textwidth]{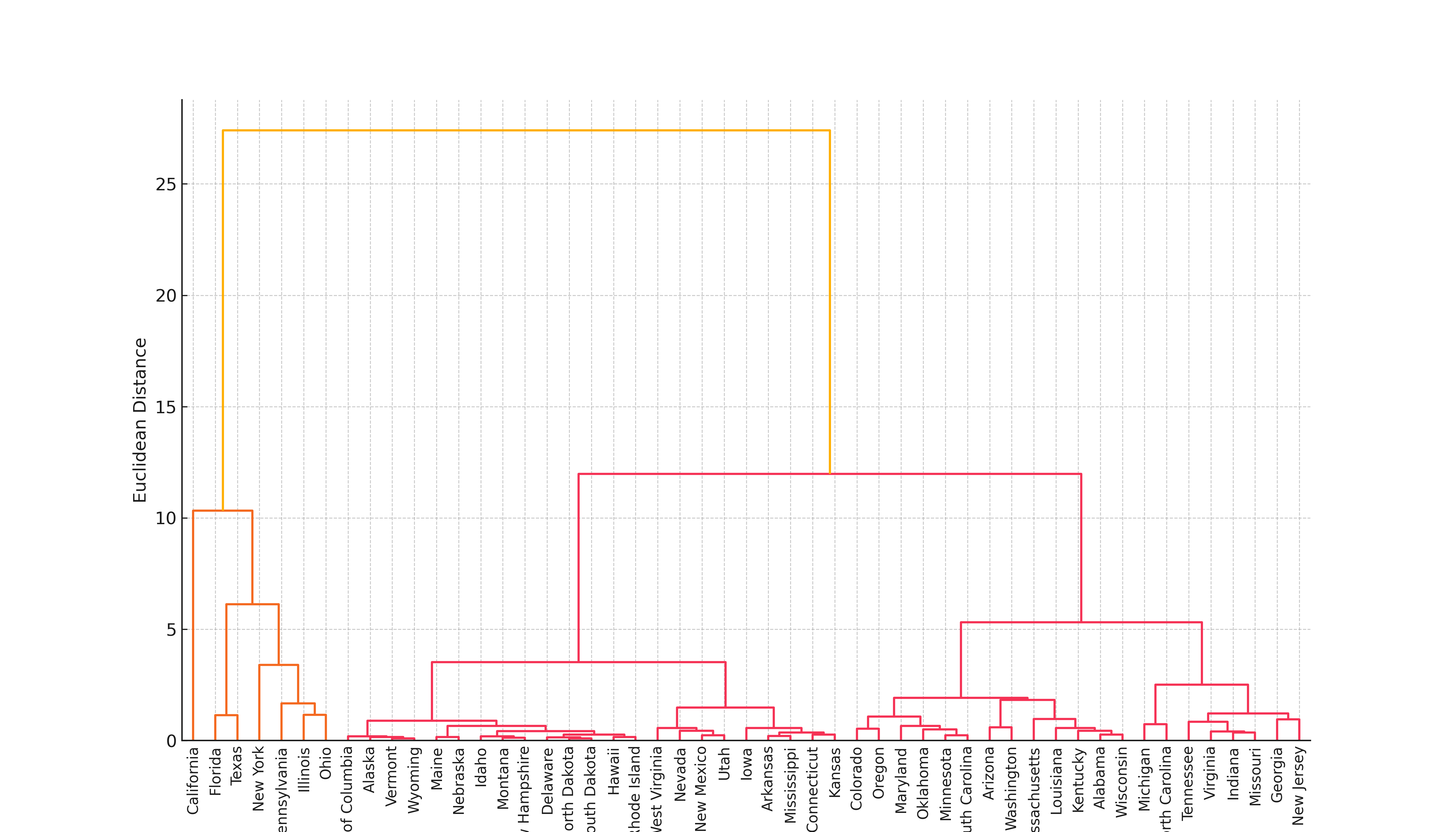}
    \caption{Hierarchical clustering dendrogram showing similarity among U.S. states based on mortality metrics. }
    \label{fig:hierarchical_clustering}
\end{figure}

As shown in Figure \ref{fig:hierarchical_clustering}, a hierarchical cluster analysis was conducted to identify structural similarities in mortality patterns among U.S. states. The dendrogram illustrates the results of an agglomerative clustering procedure based on Euclidean distance, which quantifies dissimilarity across each state’s mortality profile. States connected by shorter linkage distances exhibit greater homogeneity in age-adjusted death rates and total mortality distributions, whereas those joined at higher distances demonstrate more distinct mortality characteristics. The hierarchical structure reveals several meaningful groupings that suggest regional or demographic affinities, reflecting underlying similarities in population health, socioeconomic conditions, or healthcare access. Overall, the clustering presented in Figure \ref{fig:hierarchical_clustering} complements the preceding correlation and copula analyses by emphasizing the multidimensional dependence structure that characterizes mortality variation across the United States.

\section{Conclusion:}

This study provides a comprehensive examination of the relationship between total mortality and age-adjusted death rates using NCHS data on the leading causes of death in the United States from 2012 to 2017. Through scatterplots, boxplots, log-linear regression, empirical copula analysis, and hierarchical clustering, the results consistently demonstrate a strong positive association between crude death counts and standardized mortality risk. While heart disease and cancer dominate the upper range of both measures, the analysis reveals important nuances across causes of death, particularly for conditions that disproportionately affect younger or specific demographic subgroups. The incorporation of age adjustment proves essential for accurately interpreting mortality trends, as it separates true changes in health risk from demographic processes such as population aging. Collectively, these findings underscore the methodological importance of age standardization in public health research and highlight the value of combining descriptive and regression-based frameworks to capture both linear and nonlinear dependence in national mortality data.

At the same time, several limitations of this study suggest directions for future research. First, the analysis is restricted to a relatively short pre-pandemic window (2012–2017) and to the ten leading causes of death, which limits the ability to evaluate longer-term structural shifts, emerging causes, or the profound disruptions associated with events such as COVID-19. Second, all inferences are based on aggregate, observational data, which precludes causal interpretation and may conceal important within-group heterogeneity related to race, ethnicity, socioeconomic status, or geography. Third, the modeling framework relies on a log-linear specification and a specific empirical copula construction, both of which impose structural assumptions that may not fully capture complex tail behavior or non-monotonic dependence. Future work can address these limitations by extending the time horizon, incorporating additional causes and morbidity measures, and integrating richer covariate information through multilevel or spatial models. Moreover, comparing alternative link functions, copula families, and distributional assumptions, as well as jointly modeling mortality and morbidity outcomes, would provide a more robust and nuanced understanding of how demographic, epidemiological, and policy factors shape mortality patterns in the United States.

\section*{Declarations.}
\subsection*{Ethics approval and consent to participate.}
Not applicable.
\subsection*{Consent for publication.}
Not applicable.
\subsection*{Availability of data and material.}
Data sets  were obtained from the National Vital Statistics System (NVSS).
\subsection*{Competing interests.}
No potential conflict of interest was reported by the authors.	
\subsection*{Funding.}
Not applicable. 
\subsection*{Acknowledgements.}
 Not applicable.

\bibliographystyle{apalike}
\bibliography{bib}
\end{document}